\documentclass[conference,a4paper]{IEEEtran}
\makeatletter
\def\ps@headings{%
\def\@oddhead{\mbox{}\scriptsize\rightmark \hfil \thepage}%
\def\@evenhead{\scriptsize\thepage \hfil \leftmark\mbox{}}%
\def\@oddfoot{}%
\def\@evenfoot{}}
\makeatother
\usepackage{subfigure}
\usepackage{color}
\usepackage{balance}
\usepackage{acronym}
\usepackage{caption}

\usepackage[draft]{todonotes}

\usepackage{graphicx,times}
\usepackage[bookmarks=false]{hyperref}
            
\graphicspath{{artworks/}}
\IEEEoverridecommandlockouts

\ifCLASSINFOpdf
   \DeclareGraphicsExtensions{.pdf,.jpeg,.png}
\fi

\newcommand{\quotes}[1]{``#1''}

\begin{document}
\title{Revisiting Open eXchange Points with Software Defined Networking\vspace{-2.5ex}$^*$\thanks{*This work was partly funded by the EU G\'EANT (GN4-1) project \cite{geant}.}}

\author{\IEEEauthorblockN{
Pier Luigi Ventre\textsuperscript{\dag}, 
Bojan Jakovljevic\textsuperscript{\S},
David Schmitz\textsuperscript{\ddag},
Stefano Salsano\textsuperscript{$\gamma$},
Matteo Gerola\textsuperscript{$\tau$},
Luca Prete\textsuperscript{$\natural$},\\
Sebastiano Buscaglione\textsuperscript{$\delta$},
Jose Aznar\textsuperscript{$\epsilon$},
Kostas Stamos\textsuperscript{$\rho$}
}
\IEEEauthorblockA{
\textsuperscript{\dag}GARR/Univ. of Rome Tor Vergata, 
\textsuperscript{\S}AMRES, 
\textsuperscript{\ddag}Leibniz Supercomputing Centre, 
\textsuperscript{$\gamma$}CNIT/Univ. of Rome Tor Vergata,\\
\textsuperscript{$\tau$}CREATE-NET,
\textsuperscript{$\natural$}ONLab,
\textsuperscript{$\delta$}G\'EANT,
\textsuperscript{$\epsilon$}i2Cat,
\textsuperscript{$\rho$}GRNET
}
}


\maketitle

\begin{abstract}
The introduction of SDN in Service Providers' networks like G\'EANT is a challenging task. Prototypes of new generation services have to exhibit \quotes{carrier grade} characteristics, meet the high level expectations and specialized needs of G\'EANT customers. In this demonstration, we present the SDN based prototype of \emph{G\'EANT Open} service, used by G\'EANT customers and approved commercial partners to interconnect using Layer 2 circuits. Currently, this service is delivered through a set of Open eXchange Points leveraging on legacy solutions. The SDN based prototype has been realized on top of ONOS and leverages on hardware switches for the data plane. During the demo, which runs inside the G\'EANT Testbed Service, we show how operators can deploy services and manage the SDN based infrastructure. 
\end{abstract}

\begin{IEEEkeywords}
Software Defined Networking, Open Source, Internet eXchange Point, Open eXchange Point, Open Networking Operating System

\end{IEEEkeywords}
\vspace{-2ex}

\section{Introduction}
\vspace{-0.5ex}
\label{section:introduction}
Software Defined Networking (SDN) \cite{sdn} is a recent networking paradigm that may drastically change the way IP networks run today. However, the scientific and technological question \quotes{What is the best way to introduce SDN in IP Service Providers (ISP) networks ?} is still an open issue and different solutions have been proposed so far. G\'EANT \cite{geant}, the 500Gbps pan-European network inter-connecting 38 National Research and  Educational  Networks (NRENs), provides users and customers with a wide range of services including Point to Point circuits at all layers, optical services and IP/MPLS testbed services \cite{services}. Starting from the analysis of the G\'EANT requirements, we have realized the SDNization of the \emph{G\'EANT Open} service. Using this service, the NRENs can connect to external (non-G\'EANT) networks through the Open eXchange Points (OXP). The OXPs are similar to the standard  Internet eXchange Points (IXPs)\cite{ixp}, but with a fundamental difference: inside an OXP, the customers (NRENs or external customers) request the establishment of Layer 2 circuits between end-points.

The SDN version of the \emph{G\'EANT Open} service has been built on top of the ONOS controller \cite{onos}. ONOS is an Open Source Network Operating System (NOS) providing a logically centralized network view. For scalability, the network resources can be partitioned and controlled by different controllers. Having multiple controllers can be used also for fault tolerance. In case of failures, a backup controller gains the mastership of the switches and notifies the others instances. The data plane itself is resilient: when a link or a network device fails, ONOS automatically reroutes the traffic. At the NorthBound side, ONOS offers different abstractions to applications. One of these, are the \emph{Intents}, which provide applications with a simple and powerful abstraction of end-to-end network connectivity. Using Intents, developers can rely on high-level network services abstractions. \emph{G\'EANT Open} service is composed by an existing ONOS application (SDN-IP \cite{sdnip}) and by a new ONOS application called L2-SDX. Both are available under Apache 2.0 Open Source license. SDN-IP is shipped within ONOS, L2-SDX can be downloaded from \cite{apps}. The Intent framework has been widely used for developing the SDN-IP and L2-SDX services.

\section{Development process}
\vspace{-0.5ex}
\label{section:Development process}
\begin{figure}
\centering
\includegraphics[width=.45\textwidth]{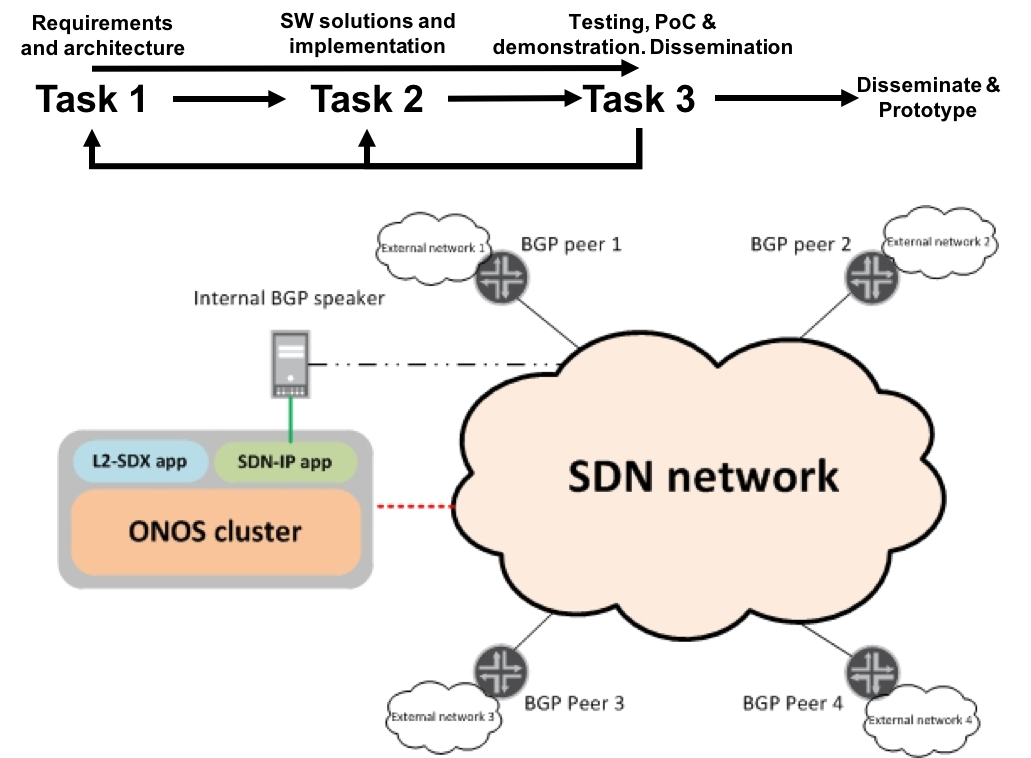}
\caption{Development process and SDN based OXP architecture}\vspace{-4ex}
\label{fig:arch}
\vspace{-2ex}
\end{figure}
The introduction of SDN in ISP networks implies finding solutions to critical requirements and issues, such as: i) how to provide high levels of scalability and fault tolerance; ii) how to guarantee reliability and resiliency in the infrastructures. In order to support both the development/testing aspects and the evaluation of different design alternatives, it is fundamental to have a well defined development process and a realistic emulating platform. As regards G\'EANT project, a controlled development process has been setup. It's efficacy has been demonstrated in GEANT 1 project and it was able to adapt to the technology changes during the years, see top of Figure \ref{fig:arch}. As regards the emulation platform, it's been possible to rely on the recent G\'EANT Testbed Service (GTS) \cite{services}. GTS delivers virtual testbeds powered by several G\'EANT Point of Presences (PoPs) co-located facilities. It offers different type of resources like VMs, SDN devices, Virtual Circuits, and interconnections with external domains through the G\'EANT transport network.
\section{SDN based OXP architecture}
\label{section:architecture}
SDN enabled devices have been adopted as data plane, while an ONOS cluster has been used to control the SDN based devices. The \emph{G\'EANT Open} components, SDN-IP and L2-SDX, run on top of ONOS, as applications. SDN-IP allows the interconnection of an SDN island with external legacy networks using the BGP protocol. L2-SDX allows customers to create Layer 2 circuits between client interfaces. L2-SDX aims to substitute \emph{G\'EANT Open} service, while SDN-IP enhances the OXPs providing IP connectivity and routing between participants through the BGP protocol. Coexistence of the services at the client interfaces is realized through the use of VLAN tags: one VLAN tag is reserved for BGP speakers (SDN-IP) and the other ones for L2-SDX users. These VLAN tags have only local significance at the client interface. The scalability is preserved as in the core different mechanisms can be used to encapsulate the service traffic.

SDN-IP allows OXP participants the exchange of IP routes and of traffic. One or more internal BGP speakers are needed to peer with the external routers and to act as bridges between the external domains and the SDN-IP application. SDN-IP has two main tasks in the SDN based infrastructure: i) to install flows for the establishment of BGP sessions; ii) to translate received routes into flows on the SDN switches, allowing the exchange of transit traffic. As regards Layer 2 circuits, the L2-SDX application provides the necessary mechanisms for the service provisioning and monitoring. Operators can manage and monitor the application through the CLI and GUI that accepts high-level customer’s requests. L2-SDX provides operators with powerful APIs and abstractions. Customers can request the provisioning of Layer 2 circuits between end-points, modeled as \emph{edge connectors}. The requests are automatically translated by ONOS into SDN flows on the devices. Moreover, it eases service management, e.g. enforcing isolation avoiding several types of conflicts: i) operators see the abstraction of managing different Virtual eXchange Points which contain a number of \emph{edge connectors}; ii) the resources (ports or VLAN tags) associated with an \emph{edge connector} cannot be reused; iii) an \emph{edge connector} can only be used in a single circuit, and iv) \emph{edge connector} in a virtual eXchange Point cannot be interconnected with a connector in another virtual eXchange Point enforcing isolation among them.

\section{Demonstration}
\label{section:demonstration}
At present, the provisioning process of the \emph{G\'EANT Open} service includes several manual operations resulting in very long provisioning times. The objective of the demonstration is to show how the introduction of SDN can improve the services life-cycle in terms of capability, flexibility and scalability.
\begin{figure}
\centering
\includegraphics[width=.48\textwidth]{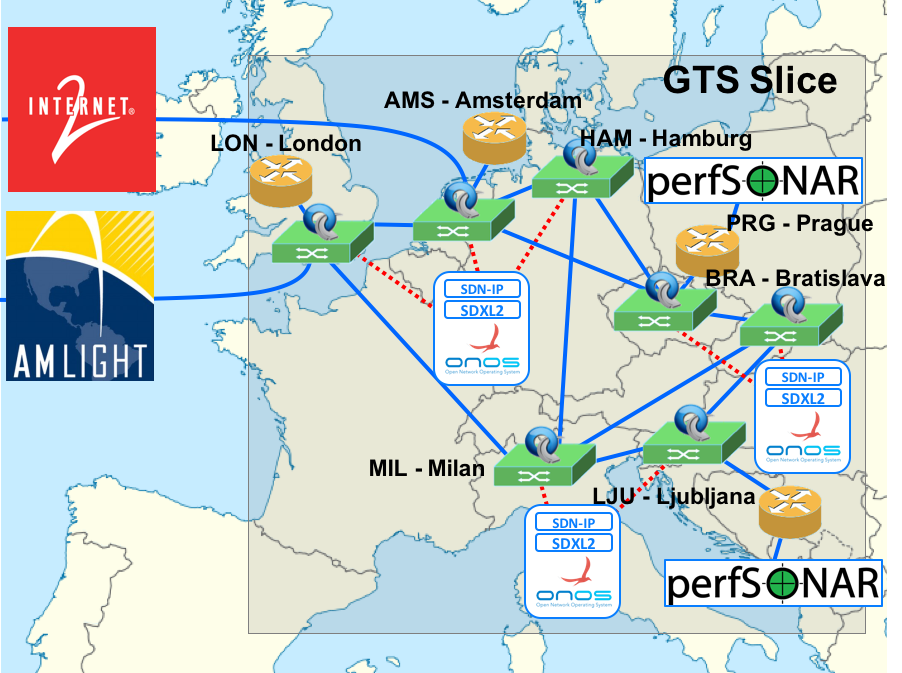}
\caption{Pilot setup}\vspace{-4ex}
\label{fig:demo}
\end{figure}
The demo runs in GTS and uses commercial hardware switches and Virtual Machines. Using GTS, we have built a large-scale PoC with 7 HP OpenFlow switches deployed in 7 G\'EANT PoPs (shown in Figure \ref{fig:demo}). The data plane is controlled by a cluster of 3 ONOS instances located in Amsterdam (AMS), Bratislava (BRA) and Milan (MIL). Four VMs, working as BGP peers and two stub networks with perfSONAR hosts have been deployed. PerfSONAR is a network performance measurement and troubleshooting tool, designed to work in multi-domain scenarios. This PoC has been integrated in a worl-wide demo hosted at Open Networking Summit 2016, where ON.Lab has successfully deployed ONOS and SDN-IP creating a global network facility entirely based on SDN.

At the beginning of the demo, the SDN-IP functionality is shown. After the activation, the application installs flows for the establishment of the BGP connections between the external domains and the internal BGP speaker located in AMS. After a while, the BGP connections are established and the IP routes are exchanged. The demonstrator shows BGP routes in the ONOS controller, the devices status and OpenFlow rules installed. Afterwards, the BGP peer located in PRG is added to the infrastructure, demonstrating the dynamic addition of a new user. The operator proves how this can be done without impacting on other users. The operator shows again the BGP routes in ONOS controller, devices status, the OpenFlow rules being installed, and demonstrates that IP connectivity between users has been established. At this point, the operator shows the L2-SDX function activating two Layer 2 circuits. The first one is between the BGP peer located in PRG and the one located in BRA. Connectivity is demonstrated through the establishment of a BGP session and a ping command. Then, the operator shows how this new addition influences the BGP protocol and SDN-IP operations which install new flows into SDN switches. This allows the transit of traffic designed for networks announced by the peer in BRA and originated from the other peers. Then, a second Virtual Circuit between the BGP peer located in LON and the one located in PRG is installed. Finally, the management functionality of L2-SDX are shown to the audience: the operator can manage the virtual eXchange Points, the \emph{edge connectors} and the Virtual Circuits. For the \emph{edge connectors} and the Virtual Circuits, operator shows also the operational status.

\vspace{-2ex}
\bibliographystyle{IEEEtran}
\bibliography{main.bbl}

\end{document}